# A preliminary study of acoustic propagation in thick foam tissue scaffolds composed of poly(lactic-co-glycolic acid)

**N G Parker [1], M L Mather[2], S P Morgan[2] and M J W Povey[1]**

[1] School of Food Science and Nutrition, University of Leeds, Leeds, LS2 9JT, UK
[2] Electrical Systems and Applied Optics Research Division, University of Nottingham, Nottingham, NG7 2RD, UK
n.g.parker@leeds.ac.uk

**Abstract**. The exclusive ability of acoustic waves to probe the structural, mechanical and fluidic properties of foams may offer novel approaches to characterise the porous scaffolds employed in tissue engineering. Motivated by this we conduct a preliminary investigation into the acoustic properties of a typical biopolymer and the feasibility of acoustic propagation within a foam scaffold thereof. Focussing on poly(lactic-co-glycolic acid), we use a pulse-echo method to determine the longitudinal speed of sound, whose temperature-dependence reveals the glass transition of the polymer. Finally, we demonstrate the first topographic and tomographic acoustic images of polymer foam tissue scaffolds.

## 1. Introduction

Tissue engineering aims at the creation of biological substitutes that replace, restore or improve tissue function [1]. Central to this is the requirement of an artificial extracellular matrix, or scaffold, on which to grow the surrogate tissue, which is either developed *in vitro* with subsequent implantation [2] or transplanted directly for *in vivo* growth [3]. Tissue engineering has demonstrated many successes so far (see [4] for a review). However, significant challenges lie ahead, notably for cell biology, biomaterials, scaffold engineering and imaging/characterization. In recent years acoustics has found diverse applications in assisting tissue engineering. Examples include the use of surface waves to promote cell proliferation in tissue scaffolds [5, 6], monitoring of cell integration [7], sound-induced acceleration of bone tissue growth [8], manipulation of scaffold structure [9], and a route to monitor scaffold formation [10] and polymer degradation [11]. Acoustic waves also offer exclusive opportunities for characterisation of scaffolds due to their ability to probe structural, mechanical and fluidic properties. It is this possibility that motivates the current work.

Of much current focus in tissue engineering are three-dimensional polymer foam scaffolds. These may be composed of a natural polymer, e.g., collagen and chitosan, or a synthetic polymer, e.g., poly(lactic acid) or poly(lactic-co-glycolic acid), all of which satisfy the rudimentary requirement of being biocompatible and biodegradable. The structural and mechanical properties of the scaffold, such as its architecture, transport networks and mechanical integrity, must be precisely controlled to cultivate the required tissue type. As a result, accurate methods to characterize the foams and assure their suitability are crucial to the advancement of this field. Furthermore, approaches to fabricating the foams often feature an inherent randomness in the final structure, e.g., supercritical $CO_2$ and salt leaching/solvent casting, accentuating the need for accurate characterization methods.

Characterisation of polymer foam scaffolds is complicated by the multitude of factors to consider. Typically, the scaffold is characterized by its structural properties, with porosity, pore shape, pore size distribution, pore interconnectivity, total pore surface and tortuosity all cited as key structural parameters for tissue cultivation (see [12] and references therein). Image-based techniques provide the leading route to determine these parameters, with micro *X*-ray CT the most widely used modality [13]. However concerns exist over the variations that arise amongst different imaging modalities, contrast and resolutions [12]. Other work has highlighted the importance of parameterising the scaffold function with dynamic/fluidic information, notably its permeability [14]. These properties require distinct measurement approaches, such as fluid porometry, which are usually invasive. Acoustics has, as yet, been untested as a route to probe the internal characteristics of polymer foam scaffolds, despite its exclusive ability to reveal the structural, mechanical and fluidic properties of foams. These capabilities are beginning to be demonstrated in other areas of tissue engineering. For example, acoustic elasticity imaging has been demonstrated in tissue-mimicking phantoms [15] and hydrogel scaffolds [16], while the mechanical integrity of titanium biomaterials have been analysed acoustically in [17]. Of noteworthy importance for acoustic characterisation techniques is Biot theory, a well-established model of acoustic propagation in fluid-saturated porous media. This model gives the potential to characterise porosity, elastic parameters, permeability [18] and average pore size [19] of foams. Indeed, this theory has proved successful in structural studies of bone [19, 20], which bears many physical similarities to foam scaffolds.

Information on the basic acoustic properties of the synthetic biopolymers is scarce. While monitoring polymer degradation, Wu *et al.* determined the speed of sound at a single temperature [11], while Mather *et al.* monitored acoustic impedance during the scaffold fabrication [10]. Meanwhile, the acoustic propagation through foam tissue scaffolds has, to our knowledge, been untested. It, for instance, remains unknown if the scaffold can be saturated sufficiently with coupling liquid to reliably support acoustic propagation, with capillary resistance, surface chemistry and disconnected regions to contend with. Our motivation here is to address these basic questions. Focussing on foam scaffolds composed of poly(lactic-co-glycolic acid) (PLGA) we first examine the acoustic properties of the bulk biopolymer. A pulse-echo analysis enables us to measure the longitudinal speed of sound over a large temperature range. Notably, the glass transition of the polymer has a marked signature in the speed of sound. We then turn to the foam structures themselves and consider the acoustic imaging of their surface properties and sub-surface structure. The ability to image an object buried within the scaffold is demonstrated, illustrating the feasibility of acoustic propagation through these structures.

## 2. Materials and methods

### 2.1. Tissue scaffolds

Tissue scaffolds were donated by the School of Pharmacy at the University of Nottingham. The scaffolds are composed of PLGA 85:15 (supplied by Lakeshore Biomaterials) with a weight average molecular weight of 53 kDa. The foam structures were generated by the supercritical $CO_2$ method [21, 22]. In brief, PLGA granules are subjected to $CO_2$ under pressure. Entering its supercritical phase the $CO_2$ diffuses through the polymer, lowering its glass transition temperature and causing it to soften. After a suitable time the pressure is vented, the $CO_2$ re-enters its gaseous phase and generates bubbles, to form the final foam. The scaffolds are rigid, opaque and white. They have approximate dimensions of 12 mm in diameter and 6 mm in height, and a mass of around 0.2g. From reference [22], where similar fabrication parameters were used, we can expect the porosity (the ratio of the volume of the pore space to the overall volume) to be around 80% and the pore sizes to lie in the range 50-250µm. Inherent to the supercritical $CO_2$ foaming method is the formation of a non-porous skin and domed roof, which we remove with a scalpel blade. Although PLGA hydrolyses in water, the degradation can be assumed to be negligible on the immersion timescales involved in this study [23].

The scaffold porosity is a key characteristic and can be derived from the expression,

$$\varepsilon = 1 - \frac{m_{\text{tot}}}{\rho_{\text{frame}} V_{\text{tot}}}. \qquad (1)$$

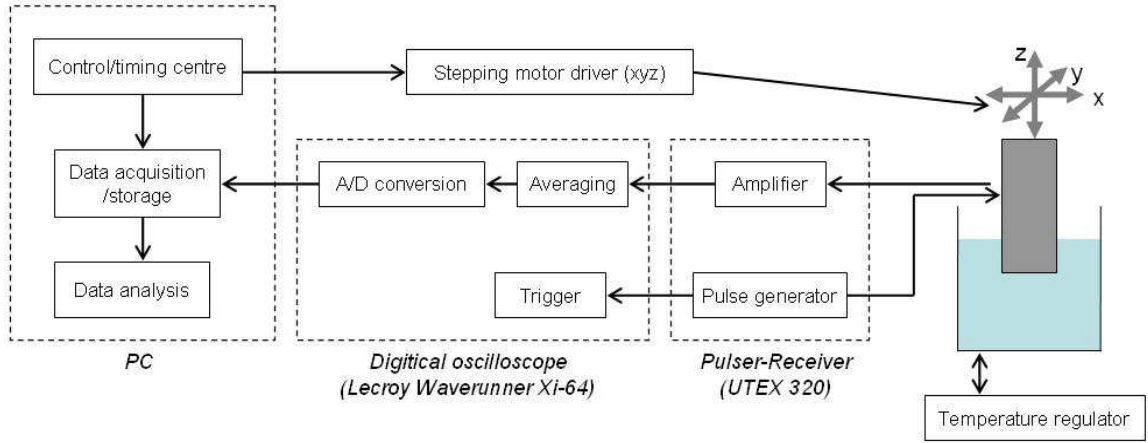

Figure 1: Block diagram of the key components within our acoustic measurement apparatus.

From mass $m_{tot}$ and volume $V_{tot}$ measurements of the scaffold (assuming the scaffold to be cylindrical), and the frame/polymer density to be *as per* the manufacturers specification of $\rho_{frame}$=1250 kg m$^{-3}$, we determine the scaffold porosity to be 81.5% (averaged over four samples).

2.2. Scanning acoustic platform
Our acoustic measurements and imaging are performed using a reflection-mode scanning acoustic platform [24]. Two ultrasound transducers are employed in this work, depending on the measurement being performed: a weakly-focussed transducer (Panametrics V311) with a centre frequency of 10 MHz and -6dB bandwidth of 5 MHz, and a tightly-focused transducer (Panametrics V3534) with an operating frequency of 55 MHz and -6dB bandwidth of 20 MHz. The generic operation of the platform is summarized in figure 1. A pulser-receiver (UTEX 320) provides a 300 V rectangular RF pulse with 10 ns duration to excite the piezoelectric transducer. The pulse repetition rate is set to 500 Hz. Acoustic reflections returning to the transducer generate RF pulses which are received and amplified by the pulser-receiver. This signal is then entered into a digital oscilloscope (Lecroy Waverunner Xi-64), where the signal is averaged over 200 sweeps to reduce noise. Following analogue-to-digital conversion the signal is exported to a PC for storage and analysis.

The acoustic platform can perform automated positioning of the transducer in three dimensions. In addition, the platform includes a sample bath which incorporates an outer fluid jacket for temperature control. A water-antifreeze mixture is circulated through this jacket by an external temperature bath/circulator (Haake B5/DC50). This enables us to regulate the temperature of the sample bath over the range 0-80$^o$C. Independent measurements of the bath temperature are performed using a high precision temperature probe (Hart Scientific 5612 probe), accurate to 0.01 $^o$C. From a previous mapping of the temperature profile within our sample unit [24], we anticipate that the variations across the probe-sample region will amount to less than 0.2$^o$C.

2.3. Acoustic measurements

*2.3.1. Sample preparation*
For measuring the acoustic properties of the polymer we require a homogeneous sample. A foam scaffold is ground into a coarse powder and placed in a cylindrical aluminium mould with diameter 4 mm and height 5 mm. The mould is heated on a hot plate to 150$^o$C for approximately 1 hour. Being above the glass transition temperature (typically 30-60$^o$C at room temperature), the polymer relaxes into the mould. Care is taken to ensure that bubbles are removed from the sample. After cooling, the end result is a transparent disc-shaped solid sample, approximately 3 mm thick and 4 mm in diameter.

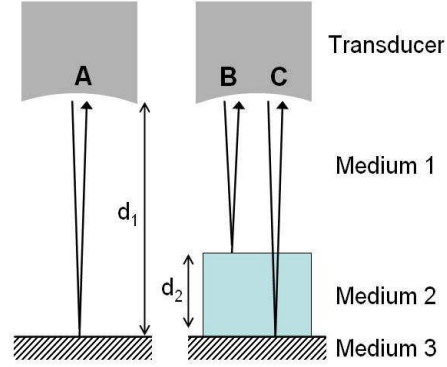

Figure 2: Schematic of the speed of sound measurements. From the time-of-flight of three sound paths (A, B and C) we determine the speed and sound in the sample.

## 2.4. Speed of sound measurements

We employ a time-of-flight technique to measure the speed of sound in this milli-scale polymer sample. Consider a homogeneous sample placed upon a substrate, immersed in a coupling fluid and insonified at normal incidence from above. From the times-of-flight of three sound paths (illustrated in figure 2) we can derive the speed of sound through the sample. Path A corresponds to the reflection from the fluid-substrate interface (in the absence of the sample). For paths B and C, the sample is present. In path B the sound pulse reflects at the first fluid-sample interface. In path C the sound pulses transmits through the sample and reflects at the sample-substrate interface. We denote the speeds of sound in the fluid and sample by $c_1$ and $c_2$, respectively, and the transducer-substrate distance and sample thickness by $d_1$ and $d_2$, respectively. The times-of-flight are then given by,

$$t_A = 2d_1/c_1 \quad (2)$$
$$t_B = 2(d_1 - d_2)/c_1 \quad (3)$$
$$t_C = 2(d_1 - d_2)/c_1 + 2d_2/c_2. \quad (4)$$

Substituting equations (1) and (2) into equation (3) and rearranging gives,

$$c_2 = c_1 \left( \frac{t_A - t_B}{t_C - t_B} \right), \quad (5)$$

from which we can readily determine the sample speed of sound $c_2$, providing the speed of sound of the coupling fluid $c_1$ is well known.

To implement this technique we insert the polymer sample and substrate (a disc of polyether ether ketone) into the sample bath of the acoustic platform, with Millipore water as the coupling fluid. The 10 MHz transducer is employed for this study. This transducer has a focal distance of $F=54$ mm and aperture diameter of $D=13$ mm, giving a semi-aperture angle of $\theta = \sin^{-1}(D/2F) = 7°$. For rays propagating at this maximal angle the acoustic path length will be larger than those propagating normal to the surface by a factor of $\sec\theta^{-1} \approx 0.007$. Since this is small we are justified in treating the sound beam as paraxial. Assuming an aqueous medium at $20°C$ our sound beam has a wavelength of $\lambda = c_1/f = 150$ μm and a pulse length of 300 μm. In the focal plane the beam diameter will be diffraction-limited to $F_r = 1.22 \, F\lambda/D = 300$μm. The depth of focus is $F_z = 8F^2 c_1/(D^2 f + 2Fc_1) = 1.7$cm. Thus, by positioning the focal plane at the substrate surface the full axial extent of the sample lies well within the focal region. The sample thickness (~3 mm) is considerably larger than both the acoustic wavelength and pulse length, ensuring the reflections are well-separated in time and are not coupled.

Each received pulse is demodulated computationally via the Hilbert transform, and its peak used to define its time-of-flight. The speed of sound in water is determined from the temperature-dependent equation of Bilaniuk and Wong [25, 26] (148-point) and the measured temperature of our sample bath.

## 2.5. Acoustic imaging of scaffolds

### 2.5.1. Scaffold preparation/permeation

For successful acoustic propagation it is essential to saturate the scaffold with fluid and remove highly-scattering gas pockets. Ideally, this fluid would be water or water-based, due to its biocompatibility and well-characterised properties. However, aqueous permeation of such scaffolds is challenging due to their hydrophobicity, tortuous pore network, capillary resistance and closed pores. To overcome hydrophobicity we use ethanol as a wetting agent, before dilution in water [27]. Scaffolds are immersed in ethanol surrounded by an ice bath. A chilled environment is necessary since ethanol lowers the glass transition temperature of the polymer from around 40$^{o}$C to 10$^{o}$C [28]. The scaffold is insonified with high intensity ultrasound throughout the permeation process (Hielscher UP100H with power output of 20 W and a duty cycle of 0.1). This is found to dramatically improve the permeation compared to immersion only. Although high intensity ultrasound has been reported to disrupt the pore structure [9], the repeatability of our measurements and absence of detectable mass loss suggest that such structural modifications are insignificant. Fluid uptake is monitored by measuring the scaffold mass at regular intervals. From this we determine that (94±9)% of the pore space becomes filled with fluid (after 2 hours). The scaffold is then immersed in Millipore water for 6 hours, during which diffusion drives the dilution of ethanol with water throughout the sample.

*2.5.2. Scanning acoustic microscopy*

Acoustic imaging of the scaffolds is performed by employing the acoustic platform as a scanning acoustic microscope. The temperature of the sample is maintained at 25$^{o}$C±0.2 $^{o}$C. We employ both the 10 MHz and 55 MHz transducers. For the former the focal region is 300 µm laterally and 17 mm axially. For the latter the focal region is 25 µm laterally and 200 µm axially. The platform performs raster-scanning of the transducer in a plane above the sample and is automated to obtain the received trace at each point. Each trace is demodulated to reveal amplitude information only. From the final data set, various imaging sections can be processed and viewed. Of these we are, firstly, concerned with topographical imaging of the scaffold surface, where the time-of-flight of the returning signal is used to determine the surface height. Secondly, we will obtain C-scan images, where the voltage emanating from a given plane/time-of-flight is viewed.

## 3. Results and discussion

3.1. Acoustic measurements

*3.1.1. Speed of sound measurements*

Typical voltage signals in the absence and presence of the solid PLGA sample are shown in figure 3(a). For this example, the ambient temperature and pulse separations are given in table 1. The thickness of the sample can be derived from the times-of-flight via $d_2=c_1(t_A-t_B)/2=2.577$ mm, in firm agreement with a calliper measurement of $d_2=2.58$ mm. From equation (4) we determine the polymer speed of sound to be $c_2=2321.8$ ms$^{-1}$. Using the quoted density of $\rho_2=1250$ kg m$^{-3}$ we calculate the acoustic impedance of PLGA to be $Z_2=\rho_2 c_2=2.90$ MRayl. The temperature of the sample bath was heated to 60$^{o}$C and swept down to 5$^{o}$C, with the pulse times recorded at regular intervals. The variation of the PLGA speed of sound with temperature is presented in figure 3(b). At low temperature (5-25$^{o}$C) the speed of sound decreases linearly at a rate of 3 m s$^{-1}$ $^{o}$C$^{-1}$. In the vicinity of 30$^{o}$C the behaviour becomes nonlinear. Beyond this the speed of sound again decreases linearly but at a greater rate of 15 m s$^{-1}$ $^{o}$C. We attribute this change to the glass transition, where the material changes from a solid to a glassy phase. Similar qualitative behaviour has been observed for polystyrene [29]. During the glassy phase pulse path C (Figure 2) becomes strongly attenuated with temperature and beyond 60$^{o}$C we no longer detect this pulse. We attribute this to the highly viscous nature of this state, which will increase the attenuation [30]. We estimate the glass transition temperature by extrapolating the linear fits in the solid and glassy regimes. From the crossing point we find the glass transition temperature of our PLGA sample to be 31.4$^{o}$C. This is somewhat lower than that observed elsewhere, e.g. around 50$^{o}$C for a PLGA 90:10 sample observed in [31], and may suggest some polymer scission is occurring during sample formation.

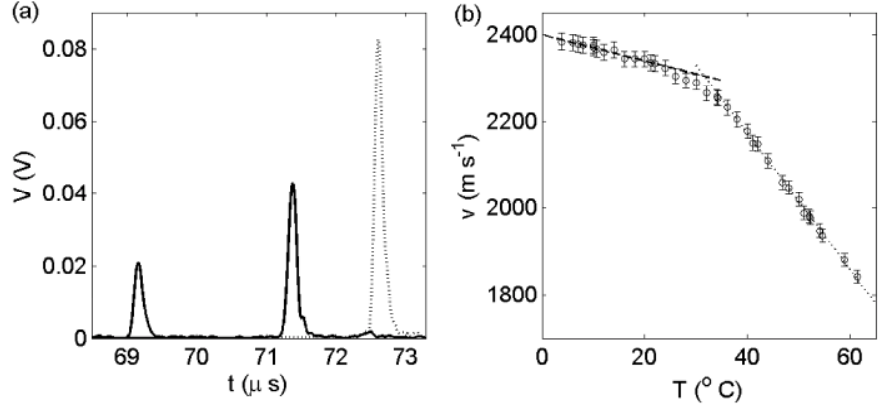

Figure 3: (a) Voltage signal in the absence (dotted line) and presence (solid line) of the PLGA sample. From left to right the pulses corresponds to paths B, C and A. (b) Speed of sound in PLGA as a function of temperature. The dashed and dotted lines are linear fits to the low and high temperature data, respectively.

| Sample | $T$ (°C) | $t_A - t_B$ (μs) | $t_C - t_B$ (μs) | $c_1$ (ms$^{-1}$) | $c_2$ (ms$^{-1}$) | $Z_2 = \rho_2 c_2$ (MRayl) |
|---|---|---|---|---|---|---|
| PLGA 85:15 | 24.02 | 3.45 | 2.22 | 1494.05 | 2321.8 | 2.90 |

Table 1: Acoustic time-of-flight measurements for the PLGA 85:15 sample.

To our knowledge, only one study has measured the speed of sound in PLGA. Wu *et al.* [11] determined a value of 2400 ms$^{-1}$ for PLGA 50:50 at 20°C. This is close to our measured value of 2344 ms$^{-1}$. Deviations can be expected due to the different polymer ratio and variations in crystallinity.

3.2. Acoustic imaging of scaffolds

*3.2.1. Scaffold permeation*

*3.2.2. Topographic acoustic imaging*

For the topographic imaging we employ the 55 MHz tightly focussed transducer due to its enhanced resolution. The top of the scaffold is aligned with the focal plane of the beam. Figure 4 shows an image of the scaffold surface, over an area of 900×700 μm. The lateral resolution is 25 μm. Vertical variations of the scaffold surface are of approximate size 500 μm. Two large pores are clearly visible at the surface, with a circular profile and diameter of approximately 200 μm. This is consistent with the study of Tai *et al.* [22] who examined similar scaffolds with *x*-ray CT and found the pore sizes to lie dominantly in the range 50-250 μm.

3.3. Tomographic imaging

Attempts to image sub-surface at 55 MHz were unsuccessful and may be due to the beam sensitivity (the tightly focussing nature the transducer makes it very sensitive to scattering and angular reflections) or high attenuation through the polymer. Subsequent imaging was performed with the 10 MHz transducer. A narrow glass slide, of height 0.5 mm and width 2 mm, was inserted transversely through the centre of the scaffold. The object was approximately 3 mm below the surface of the scaffold. The focal plane of the transducer was aligned with the midpoint of the scaffold. Due to the large focal depth the whole axial extent of the scaffold remained in focus.

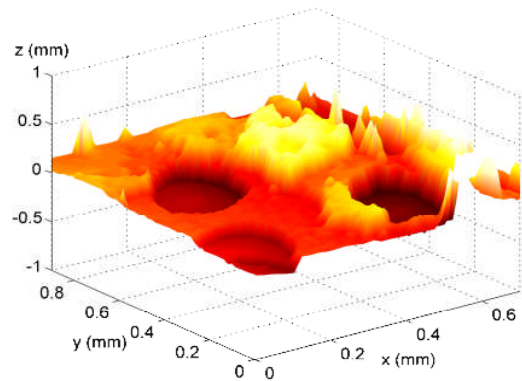

Figure 4: Acoustic image of the scaffold surface, obtained using the 55 MHz transducer. The colour indicates the vertical position, with light/dark regions representing high/low height, respectively.

Various *C*-scans are presented in figure 5(a). The first *C*-scan shows the scaffold surface. While the beam resolution (300 μm) is insufficient to spatially resolve the pores a strongly fluctuating signal indicates the uneven surface. To help identify internal features in the scaffold it is useful to consider a plot the maximum voltage received as a function of depth (figure 5(b)). Beyond the scaffold surface this voltage decays roughly with distance due to the compounded reflections that occur through the foam structure. However, we also see occasional peaks in the voltage. The peak labelled (ii) arises from the presence of a trapped air pocket, as shown in the *C*-scan in figure 5(a)(ii). (A similar bubble is responsible for the peak at $z\sim 52$ mm.) The peak labelled (iii) arises from the inserted glass slide. The rectangular shape of the slide is clearly apparent in the corresponding *C*-scan [figure 5(a)(iii)]. Note that this image is contorted due to the distortion of the beam through the scaffold. Finally, the peak voltage increases at point (iv), due to reflections from the underlying substrate. This is further confirmed from the *C*-scan in figure 5(a)(iv): the substrate surfaces appears as an arc of high voltage due to a slight angle between the surface and the *x-y* plane. Additionally, an acoustic shadow is evident in line with the slide position. These results clearly illustrate that acoustic propagation, and indeed imaging, is possible through several millimetres of hydrated foam tissue scaffold.

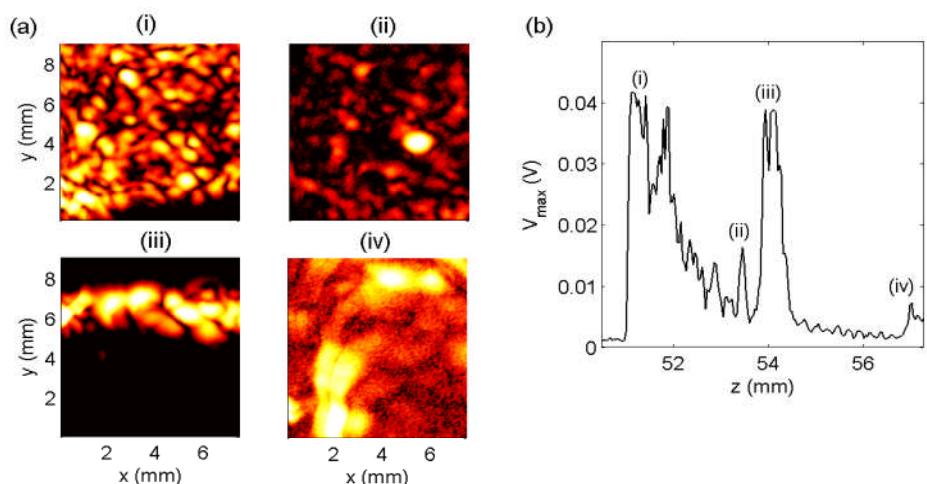

Figure 5: (a) C-scan images, taken using the 10 MHz transducer, at varying depths through the scaffold showing (i) the scaffold surface, (ii) a trapped air pocket, (iii) the embedded glass slide, and (iv) the underlying substrate (which appears as an arc of high intensity due to it being slightly tilted). The focal plane of the beam is aligned with the scaffold midpoint. (b) The maximum voltage in the C-scan as a function of depth, with scans (i-iv) indicated.

## 4. Conclusions

We have made a preliminary study into the acoustic properties and acoustic propagation in foam tissue scaffolds composed of poly(lactic-co-glycolic acid).  Using a time-of-flight analysis we measured the longitudinal speed of sound in the bulk polymer over the temperature range 5-60$^\circ$C. The marked change in speed of sound at glass transition of the polymer enables us to estimate the glass transition temperature to be 31.4$^\circ$C.  This sensitivity to phase, as well as the sensitivity to biodegradation observed elsewhere [11], highlight the potential for acoustic characterisation of the bulk polymer properties.. Furthermore, knowledge of these parameters is essential for the purposes of acoustically imaging and characterising such scaffolds.

Turning to the foam scaffolds themselves, we demonstrate that the scaffold can indeed be saturated sufficiently with water to support acoustic propagation. We employ an efficient protocol to achieving this involving pre-wetting with ethanol to overcome hydrophobicity and high intensity ultrasound to promote diffusion.  Note that certain fabrication techniques feature a stage in which the scaffold is fluid-saturated, e.g. solvent casting/salt leaching and emulsification techniques.  We then performed acoustic imaging of the scaffold surface and sub-surface.  Our surface images demonstrate strong imaging contrast of the polymer and the ability to resolve the surface and pore morphology. Meanwhile, our imaging of an object buried several millimetres into the scaffold demonstrates the ability for acoustic waves to propagate through the scaffolds and return spatial information.  The possibilities for acoustic scaffold characterization will be pursued in future work.  Finally it should be noted that acoustic propagation through foam scaffolds may aid in the functional imaging of tissue development during cultivation.  Conventional approaches based on optics fail in thick samples due to the highly scattering nature of both the scaffolds and the growing tissue.  The much weaker scattering experienced by acoustic waves may be exploited to enable deep functional imaging through the use of ultrasound contrast agents or the hybrid techniques of ultrasound-modulated optical propagation and photo acoustics.

**Acknowledgements**
We thank Dr Lisa White (Tissue Engineering Group, School of Pharmacy, University of Nottingham) for providing the tissue scaffolds used in this work, and the Biotechnology and Biology Science Research Council for funding (BBSRC ref: BB/F004923/1).